\documentclass[12pt,a4paper]{article}
\usepackage{amsmath,amsfonts,amssymb,amsthm}
\usepackage{fullpage}

\title{Spin-Hall effect with quantum group symmetries}

\author{Giovanni Landi \ 
\\[12pt]
Dipartimento di Matematica e Informatica,
      Universit\`a di Trieste,\\
      Via A. Valerio 12/1, 34127 Trieste \\ 
and INFN, Sezione di Napoli, Napoli, Italy \\
\texttt{landi@univ.trieste.it}
}
\date{January 10th, 2006}

\newcommand{\half}{\frac{1}{2}}
\newcommand{\be}{\begin{equation}}
\newcommand{\bea}{\begin{eqnarray}}
\newcommand{\ee}{\end{equation}}
\newcommand{\eea}{\end{eqnarray}}
\newcommand{\bean}{\begin{eqnarray*}}
\newcommand{\eean}{\end{eqnarray*}}
\newcommand{\nn}{\nonumber}

\newcommand{\ii}{{\,{\rm i}\,}}
\newtheorem{theo}{Theorem}[section]
\newtheorem{prop}[theo]{Proposition}

\newtheorem{defi}[theo]{Definition}
\newtheorem{rema}[theo]{\bf Remark}
\newtheorem{exam}[theo]{\bf Example}
\newcommand{\btheo}{\begin{theo} ~~\\}
\newcommand{\etheo}{\hfill $\Box$  \end{theo} }
\newcommand{\bprop}{\begin{prop} ~~\\ }
\newcommand{\eprop}{\hfill $\Box$ \end{prop} }
\newcommand{\bdefi}{\begin{defi} ~~\\ }
\newcommand{\edefi}{\hfill $\Box$ \end{defi} }
\newcommand{\brema}{\begin{rema} ~~\\ }
\newcommand{\erema}{ \hfill $\Box$ \end{rema} }
\newcommand{\bexam}{\begin{exam} ~~\\ }
\newcommand{\eexam}{\hfill $\Box$ \end{exam} }

\newcommand{\ch}{{\mathcal H}}

\newcommand{\cs}{{\mathcal S}}

\newcommand{\IC}{{\mathbb C}}

\newcommand{\II}{{\mathbb I}}

\newcommand{\IP}{{\mathbb P}}

\newcommand{\IR}{{\mathbb R}}

\newcommand{\IT}{{\mathbb T}}

\def\into{\hookrightarrow}

\newcommand{\wt}{\widetilde}
\def\bar#1{\overline{#1}}

\newcommand{\ot}{\otimes}

\newcommand{\op}{\oplus}
\DeclareMathOperator{\SU}{SU}

\DeclareMathOperator{\U}{U} 
\DeclareMathOperator{\SO}{SO}
\DeclareMathOperator{\Spin}{Spin}

\newbox\ncintdbox \newbox\ncinttbox
\setbox0=\hbox{$-$} \setbox2=\hbox{$\displaystyle\int$}
\setbox\ncintdbox=\hbox{\rlap{\hbox to
\wd2{\hskip-.125em\box2\relax \hfil}}\box0\kern.1em}
\setbox0=\hbox{$\vcenter{\hrule width 4pt}$}
\setbox2=\hbox{$\textstyle\int$}
\setbox\ncinttbox=\hbox{\rlap{\hbox to
\wd2{\hskip-.175em\box2\relax \hfil}}\box0\kern.1em}

\def\R{\IR_\theta}
\def\S{S_\theta}
\newcommand{\Sk}{S_{\theta'}}
\newcommand{\Ort}{\SO_\theta(5)}
\newcommand{\Orb}{\SO_\theta(4)}
\newcommand{\oenv}{U(so(5))}
\newcommand{\env}{U_\theta(so(5))}
\newcommand{\enor}{U_\theta(so(4))}
\def\dd{\mathrm{d}}

\begin{document}

\maketitle

\begin{abstract}
We  construct a model of spin-Hall effect on a noncommutative four sphere $\S^4$ with isospin 
degrees of freedom, coming from a noncommutative instanton, and invariance under a quantum group $\Ort$. 
The corresponding representation theory  allows to explicitly diagonalize the Hamiltonian and construct the ground state; there are both integer and fractional excitations. Similar models exist  on  higher dimensional spheres $S_\Theta^N$ and projective spaces $\IC\IP_\Theta^N$.
\end{abstract}

\vskip 3cm

Dedicated to Rafael Sorkin with friendship and respect.

\vfill

{\parindent=0pt 
\textit{Key words and phrases}: Higher dimensional Hall effects, integer and fractional exitations, quantum groups, noncommutative gauge theories, noncommutative instantons. \\ 

}

\thispagestyle{empty}

\newpage
\section{Introduction}

The Laughlin wave functions for the fractional quantum Hall effect \cite{L83} is not translationally invariant. 
This problem was overcome in \cite{H83} by a model on a sphere with a magnetic monopole at the origin. The full Euclidean group of symmetries of the plane is recovered from the rotation group $\SO(3)$ of symmetries of the sphere. From a more mathematical point of view what one is considering is the Hopf fibration of the sphere $S^3$ over the sphere $S^2$ with $\U(1)$ as 
gauge (or structure) group. 

The next Hopf fibration of the sphere $S^7$ over the sphere $S^4$, with $\SU(2)$ as gauge group and $\SO(5)$ as symmetry group, was used in \cite{ZH01} to construct a generalization of the Hall effect with (iso-)spin degrees of freedom coming from 
a $\SU(2)$ Yang monopole \cite{Y78} (the same as the $\SU(2)$ instanton of \cite{BPST}).
After that, there has been intense theoretical work  on generalizations of the Hall effect with 
the construction of models in higher dimensions and with corresponding internal degrees of freedom
(see for instance \cite{KN02,EP02,BHTZ03,HK04,BZ05,M05}).

While is not clear if these models yield theories which are liable to describe the so called spin-Hall effect 
for which experimental observations have been recently reported \cite{K04,W04},
they are nonetheless interesting and can be studied on their own. On the one hand they exhibit interesting geometrical structures (see for instance \cite{S02,M03}). 
On the other hand they have been studied in connection with topological field theories, string theory and matrix models (see for instance \cite{F02,C02,Z02}). In this context methods of noncommutative geometry (or more precisely of fuzzy geometry) are used to construct and analyze these models. 

We propose novel models of spin-Hall effect with are invariant under the action of a quantum group. 
The model that we work out explicitly here is on a noncommutative four sphere $\S^4$ (or an associated 
noncommutative plane $\R^4$), $\theta$ being a deformation parameter, with an $\SU(2)$ noncommutative instanton and invariance under the quantum orthogonal group $\Ort$. Orbital 
symmetries make up a quantum $\Orb$.
Given the many symmetries, the Hamiltonian describing the excitations of the `electron gas'  is diagonalized explicitly giving both the energies and the eigenstates of the system. There are integer as well as fractional excitations.  At the classical value of the deformation parameter, $\theta=0$, we recover fully the model of \cite{ZH01}. 

As will be reported elsewhere, similar 
 models can be defined on more general noncommutative manifolds, notably spheres $S^N_\Theta$ (and 
planes $\IR^N_\Theta$) and projective spaces $\IC\IP_\Theta^N$; here $\Theta$ is a real antisymmetric matrix of deformation parameters.  These noncommutative manifolds which are intimately related to the noncommutative tori $\IT_\Theta^N$ were introduced in \cite{CL01} and have a rich geometrical content; it is worth stressing that they are not fuzzy spaces. 
The 2 dimensional  noncommutative torus and techniques from noncommutative geometry 
played a crucial role in the analysis of \cite{BvES94} of the integer quantum Hall effect. For a noncommutative approach to the fractional quantum Hall effect we refer to \cite{MM05}.

\section{Noncommutative spheres}
 
Toric  noncommutative manifolds $M_\Theta$ were constructed and studied in \cite{CL01}. One starts with any (Riemannian spin) manifold $M$ carrying a torus action and then deforms the torus to a noncommutative one governed by a real antisymmetric matrix $\Theta$ of deformation parameters. These noncommutative manifolds were indeed named isospectral deformations in that they can be endowed 
with the structure of a noncommutative Riemannian spin manifold via a spectral triple $(C^\infty(M_\theta), D, \ch$) with the properties of \cite{C96}. 
For this class of examples, the Dirac operator $D$ is the classical one and $\ch=L^2(M,\cs)$ is the usual Hilbert space of spinors 
 on which the algebra $C^\infty(M_\theta)$ acts in a twisted manner. Thus one twists the algebra and its representation while keeping the geometry unchanged.  The resulting noncommutative geometry is 
 isospectral and all spectral properties are preserved including the dimension. Both the algebra and its action on spinors can be given via a ``star-type" or ``Moyal-type" product. The starting example of \cite{CL01}, the archetype of all these deformations, was a four dimensional sphere $\S^4$. 
Noncommutative spheres $S^N_\Theta$ in any dimensions were made more explicit in \cite{CD02}, and one can also explicitly describe projective spaces $\IC\IP_\Theta^N$. 
 
In the present paper we shall use the sphere $\S^4$ and an  $\SU(2)$ noncommutative principal fibration over $\S^4$ constructed in \cite{LS04}. Similar fibrations can be constructed on higher dimensional noncommutative spheres and also on noncommutative projective spaces.

With $\theta$ a real parameter, the algebra $A(\S^4)$ of polynomial functions on the sphere $\S^4$ is 
 generated by elements  $z_0=z_0^*, z_j, z_j^*$, $j=1,2$, subject to relations
\be\label{s4t}
z_\mu z_\nu = \lambda_{\mu\nu} z_\nu z_\mu, \quad  z_\mu z_\nu^* = \lambda_{\nu\mu} z_\nu z_\mu^*,
\quad z_\mu^* z_\nu^* = \lambda_{\mu\nu} z_\nu^* z_\mu^*, \quad \mu,\nu = 0,1,2 ,
\ee
with deformation parameters given by
\be
\lambda_{1 2} = \bar{\lambda}_{2 1} =: \lambda=e^{2\pi \ii \theta}, 
\quad \lambda_{j 0} = \lambda_{0 j } = 1, \quad j=1,2 ,
\ee
 and toghether with the spherical relation $\sum_\mu z_\mu^* z_\mu=1$. For $\theta=0$ one recovers 
the 
$*$-algebra of complex polynomial functions on the usual sphere $S^4$.

As mentioned above, the sphere $\S^4$ comes with the structure of a noncommutative Riemannian 
spin manifold (a spectral triple) $(C^\infty(\S^4), D, \ch$) with the undeformed Dirac operator $D$ on $\ch=L^2(S^4,\cs)$, the undeformed Hilbert space of spinors,  
 on which the algebra $C^\infty(\S^4)$ acts in a twisted manner. The noncommutative geometry being
 isospectral it follows that $\S^4$ is four dimensional. On $\S^4$ there is a compatible exterior calculus: 
 forms are generated by the elements $z_\mu, z_\nu^*$ in degree $0$ and elements $\dd z_\mu,\dd 
z_\nu^*$ of degree $1$ with relations
\bea \label{rel:diff}
 &\dd z_\mu \dd z_\nu+ \lambda_{\mu\nu} \dd z_\nu \dd z_\mu =0 , \quad \dd z_\mu^* \dd z_\nu+ 
 \lambda_{\nu\mu} \dd z_\nu \dd z_\mu^* =0, \quad \dd z_\mu^* \dd z_\nu^*+ \lambda_{\mu\nu} \dd z_\nu^* 
 \dd z_\mu^* =0, 
 & \nn  \\ &z_\mu \dd z_\nu = \lambda_{\mu\nu} \dd z_\nu z_\mu, \quad z_\mu^* \dd z_\nu = 
\lambda_{\nu\mu} \dd z_\nu z_\mu^*, \quad z_\mu^* \dd z_\nu^* = \lambda_{\mu\nu} \dd z_\nu^* z_\mu^*.&
\eea
 Moreover, $(\dd \omega)^*=\dd \omega^*$ and $(\omega_1\omega_2)^* = (-1)^{d_1 d_2}\omega_2^* 
\omega_1^*$ for $\omega_j$ a form of degree $d_j$.

\medskip

The noncommutative sphere $\S^4$ can be viewed \cite{CD02} as the 
 ``one-point compactification" of a noncommutative 4-plane $\R^4$. 
With $ |\zeta|^2 = \sum_{j=1}^2 z_j^* z_j$, 
the algebra $A(\S^4)$ is slightly enlarged
by a hermitian central generator $(1+| \zeta 
|^2)^{-1}$ with relations 
\[
(1+\sum_{j=1}^2 z_j^* z_j) (1+| \zeta |^2)^{-1} = 1, \qquad (1+| \zeta |^2)^{-1} (1+\sum_{j=1}^2 z_j^* z_j) = 1 . 
\]
The elements
\be\label{chart}
\wt{z}_0 = \frac{1-| \zeta |^2}{1+| \zeta |^2} \, , \quad \wt{z}_j = \frac{2z_j}{1+| \zeta |^2} \, ,
\quad \wt{z}_j^* = \frac{2z_j^*}{1+| \zeta |^2} \, , 
\ee
satisfy the same relations as the elements $z_0, z_j, z_j^*$. The difference is that the classical 
point $z_0=-1, z_j=z_j^*=0$ of $\S^4$ is not in the spectrum of $\wt{z}_0, \wt{z}_j, \wt{z}_j^*$ and 
 we are describing a noncommutative $\R^4$ whose generators can be taken to be $\wt{z}_j, \wt{z}_j^*$, 
 $j=1,2$, with relations deduced from \eqref{s4t}. In fact, the element $(1+| \zeta |^2)^{-1}$ is 
 smooth and one can deal with the algebra $C^\infty(\R^4)$ of smooth functions on $\R^4$. One can 
then 
 cover $\S^4$ with two ``charts'' with domain the noncommutative $\R^4$ and transitions on $\R^4 
\setminus \{0\}$, the point $z_j=z_j^*=0$ being a classical point  of $\R^4$. 

\section{A noncommutative Hopf fibration}
 The sphere $\S^4$ comes with a noncommutative vector bundles endowed with a self-dual gauge connection 
 \cite{CL01} which is the generalization of the BPST instanton \cite{BPST} of $\SU(2)$ Yang-Mills 
 theory. This configuration, originally defined on $\IR^4$, when conformally mapped to $S^4$ 
 coincides with the monopole configuration found by Yang \cite{Y78}. We shall describe at 
 length the noncommutative instanton connection later on since it plays a central role in the spin-Hall effect that we are 
going to introduce.

Let us start with the $\SU(2)$ noncommutative principal fibration $\Sk^7 \to \S^4$ constructed in \cite{LS04}. 
 With $\lambda'_{a b} = e^{2 \pi \ii \theta'_{ab}}$ and $(\theta'_{ab})$ a real antisymmetric matrix, 
 the algebra $A(\Sk^7)$ of polynomial functions on the sphere $\Sk^7$ is generated by elements  
$\psi_a, \psi_a^*$, $a=1,\dots,4$, subject to relations
\be\label{s7t}
\psi_a \psi_b = \lambda_{a b} \psi_b \psi_a, \quad  \psi_a \psi_b^* = \lambda_{b a} \psi_b^* \psi_a,
\quad \psi_a^*\psi_a^* = \lambda_{a b} \psi_b^* \psi_a^* ,
\ee
and with the spherical relation $\sum_a \psi_a^* \psi_a=1$. 
At $\theta=0$, it is the $*$-algebra of 
 complex polynomial functions on the sphere $S^7$. Again there is a noncommutative geometry via a  
spectral triple which is isospectral and a compatible exterior calculus. 
 As before, forms are generated by the elements $\psi_a, \psi_b^*$ in degree $0$ and elements $\dd 
\psi_a,\dd \psi_b^*$ of degree $1$ with relations similar to the ones in \eqref{rel:diff}.

 In order to construct the noncommutative Hopf bundle over the given 4-sphere $\S^4$, we need to select 
 a particular noncommutative 7 dimensional sphere $\Sk^7$. We take the one corresponding to the 
following deformation parameters
\be\label{lambda7}
\lambda'_{ab}= 
\begin{pmatrix} 1 & 1 & \bar{\mu} & \mu \\ 
1 & 1 & \mu & \bar{\mu} \\
\mu & \bar{\mu} &1 & 1\\ 
\bar{\mu} & \mu &1 & 1 
\end{pmatrix}, \quad \mu = \sqrt{\lambda}, \qquad \mathrm{or} \qquad
\theta'_{ab}=\frac{\theta}{2}\begin{pmatrix} 0 & 0 & -1 & 1 \\ 
0 & 0 & 1 & -1 \\
1 & -1 & 0 & 0 \\ 
-1 & 1 & 0 & 0  \end{pmatrix}.
\ee
The previous choice is essentially the only one that  allows the algebra $A(\Sk^7)$ to carry an action 
 of the group $\SU(2)$ by automorphisms and such that the invariant subalgebra coincides with 
 $A(\S^4)$.  The best way to see this is by means of the matrix-valued function on $A(\Sk^7)$ (we are 
changing notations with respect to \cite{LS04}) 
\be\label{Psi} 
\Psi  =
\begin{pmatrix} 
\psi_1 & - \psi^*_2 \\ 
\psi_2 & \psi^*_1 \\
\psi_3 & -\psi^*_4 \\
\psi_4& \psi^*_3
\end{pmatrix}.
\ee
Then, the commutation relations of the algebra $A(\Sk^7)$, with deformation parameter in \eqref{lambda7}, gives that  $\Psi^\dagger \Psi = \II_2 $. As a consequence, the matrix-valued function $p = \Psi \Psi^\dagger$ is a projection,  $p^2=p=p^\dagger$, and its entries rather that functions in $A(\Sk^7)$ are (the generating) elements of $A(\S^4)$. Indeed, the right action of $\SU(2)$ on $A(\Sk^7)$ is simply given by
\be \label{actionSU2}
\alpha_w (\Psi) = \Psi w , \qquad w = \begin{pmatrix} w_1 & -\bar{w}_2 \\ w_2 & \bar{w}_1 \end{pmatrix} \in \SU(2), 
\ee
from which the invariance of the entries of $p$ follows at once: $p_w = \alpha_w (\Psi)\alpha_w (\Psi)^\dagger = p$. Explicitly one finds that,  
\be \label{proj}
p= \half \begin{pmatrix}
1+z_0 & 0 & z_1 & - \bar{\mu} z_2^* \\
0 & 1+z_0 & z_2  & \mu z_1^* \\
z_1^*& z_2^* & 1-z_0 & 0\\
-\mu z_2 & \bar{\mu}   z_1 & 0 & 1-z_0 
\end{pmatrix}, 
\ee
with the generators of $A(\S^4)$ identified as bilinears in the $\psi,\psi^*$'s and given by
\bea \label{sub}
z_1 &=& 2 (\mu \psi^*_3 \psi_1 + \psi^*_2 \psi_4), \qquad
z_2 = 2(- \psi^*_1 \psi_4 + \bar{\mu} \psi^*_3 \psi_2),  \nn \\
 z_0 &=& \psi^*_1 \psi_1 + \psi^*_2 \psi_2 - \psi^*_3 \psi_3 - \psi^*_4 \psi_4 \nn \\
 &=& 2(\psi^*_1 \psi_1 + \psi^*_2 \psi_2) -1 = 1 - 2(\psi^*_3 \psi_3 + \psi^*_4 \psi_4). 
\eea
By using the commutation relations of the $\psi$'s, one straightforwardly computes that $z_1^* z_1 + z_1^* z_1  + z_0 ^2 = 1$ and  the commutation rules 
$z_1 z_2  = \lambda z_2 z_1$, $z_1 z_2^* = \bar{\lambda} z_2^* z_1$, and that  $z_0$ is central.
The relations \eqref{sub} could be expressed in the form 
\be
z_\mu = \sum_{ab}\psi_a^*(\Gamma_\mu)_{ab}\psi_b ,
\ee  
with $\Gamma_\mu$ twisted 4 by 4 Dirac matrices,
\be
\label{tga}
\Gamma_0 = \left( \begin{smallmatrix}
1  & & & \\
 & 1 &  &   \\
 &   & -1 &  \\
 &  & & -1 
\end{smallmatrix} \right), 
\qquad 
\Gamma_1 =\begin{pmatrix} 
0 & \begin{smallmatrix} 0 & 0 \\ 0 & 1 \end{smallmatrix} \\ 
\begin{smallmatrix} \mu & 0 \\ 0 & 0 \end{smallmatrix} & 0 \\
\end{pmatrix}, 
\qquad 
\Gamma_2 =\begin{pmatrix} 
0 & \begin{smallmatrix} 0 & -1 \\ 0 & 0 \end{smallmatrix} \\ 
\begin{smallmatrix} 0 & \bar{\mu} \\ 0 & 0 \end{smallmatrix} & 0 \\
\end{pmatrix},  
\ee 
and $\Gamma_{\bar{1}}=\Gamma_1^*$, $\Gamma_{\bar{2}}=\Gamma_2^*$.

\medskip
The precise sense in which the algebra inclusion  $A(\S^4) \into A(\Sk^7)$  is a (nontrivial) 
noncommutative $\SU(2)$ principal bundle is explained in \cite{LS04}.  The projection $p$ determines a 
noncommutative vector bundle $E$ over $\S^4$ by giving the collection of  its sections (matter fields) as $\Gamma(\S^4, 
E)=p [A(\S^4)]^4$
which is technically a right $A(\S^4)$-module (i.e. one can multiply a  section by a function on the 
right; not on 
the left, due to noncommutativity). On this bundle one defines a gauge  connection by $\nabla = p \circ 
\dd$ with topological charge equal to -1 and whose curvature $\Omega=p (\dd p)^2$ satisfies an 
anti-selfdual equation $\ast_\theta \Omega = -\Omega$; here the Hodge  operator $\ast_\theta$ is 
defined using the metric of the Dirac operator on $\S^4$ \cite{CD02}.  
With $su(2)$ the Lie algebra of $\SU(2)$, the $su(2)$-valued connection 1 
form (gauge potential) on $\Sk^7$ is most simply written in terms of the  matrix-valued function $\Psi$ 
in \eqref{Psi}: it is given by
\be\label{co1fo}
\omega = \Psi^\dagger \dd \Psi \, .
\ee
We shall presently ``pull it back'' to an $su(2)$-valued gauge potential on $\S^4$, in fact to ``local'' expressions on the local chart $\R^4$. Classically ({i.e.} when $\theta =0$) one obtains the charge -1 instanton of Yang-Mills gauge theory. The construction of noncommutative instantons will be described at length in \cite{LS05}. 

\section{Gauge fields over $\S^4$}
We shall first introduce a ``local section'' of the principal bundle  $\Sk^7 \to \S^4$ on the local 
chart of $\S^4$ defined in \eqref{chart}. Let $u=(u_1,u_2)$ be a complex spinor of modulus one, $u_1^* 
u_1 + u_2^*u_2 = 1$, and define
\be \label{lose}
\begin{pmatrix}
\psi_1 \\ \psi_2  
\end{pmatrix} = \rho 
\begin{pmatrix}
u_1 \\ u_2  
\end{pmatrix}
, \qquad 
\begin{pmatrix}
\psi_3 \\ \psi_4  
\end{pmatrix} = \rho \begin{pmatrix}
z_1^*& z_2^* \\ -\mu z_2 & \bar{\mu}   z_1 
\end{pmatrix} 
\begin{pmatrix}
u_1 \\ u_2  
\end{pmatrix}.
\ee
Here $\rho$ is a central element such that $\rho^2 = (1+| \zeta |^2)^{-1}$,  with $| \zeta |^2 = z_1^* z_1 + 
z_2^* z_2$ as in \eqref{chart} and the commutations rules of  the $u_j$'s with the $z_k$'s are 
dictated 
by those of the $\psi_j$: 
\be
u_1 z_j = \mu z_j u_1 \, , \quad u_2 z_j = \bar{\mu} z_j u_2 \, , \quad j=1,2 \, . 
\ee
The right action of $\SU(2)$ rotates the vector $u$ while mapping to  the ``same point" of $\S^4$, 
which, from the choice in \eqref{lose} is found to be
\be 2 (\mu \psi^*_3 \psi_1 + \psi^*_2 \psi_4) = \wt{z}_1, \,  \quad 2(- \psi^*_1 \psi_4 + \bar{\mu} 
\psi^*_3 \psi_2) = \wt{z}_2, \, \quad 2(\psi^*_1 \psi_1 + \psi^*_2 \psi_2) -1 = \wt{z}_0,
\ee 
and is in the local chart \eqref{chart}, as expected. 

By writing the unit vector $u$ as an $\SU(2)$ matrix,  $u=\left(\begin{smallmatrix} u_1 & -u_2^* \\ u_2 
&   u_1^*\end{smallmatrix}\right)$, and by substituting  \eqref{lose} into \eqref{co1fo}, we get the 
gauge potential 1-form  on $\S^4$ as 
\be
\Psi^\dagger \dd \Psi=\eta_{(u)},
\ee
 with 
\bea\label{gpot}
u \eta_{(u)} u^* + u \dd u^* &=& \frac{1}{1+| \zeta |^2} \Big( (z_1 \dd z_1^* - z_1^* \dd z_1 - 
z_2^* \dd z_2 + z_2^* \dd z_2 ) \sigma_3 \nn \\
&& \qquad \qquad + 2 (z_1 \dd z_2^* - \bar{\lambda} z_2^* \dd z_1) \sigma_+ + 
2 (z_2 \dd z_1^* - \lambda z_1^* \dd z_2) \sigma_- \Big) \, .
\eea
The corresponding gauge curvature (field strength) is found to be
\be\label{gcur}
u F_{(u)} u^* = \frac{1}{(1+| \zeta |^2)^2} \Big( (\dd z_1 \dd z_1^* - \dd z_2 \dd z_2^*) \sigma_3  
+ 2 (\dd z_1 \dd z_2^*) \sigma_+ + 2(\dd z_2 \dd z_1^* ) \sigma_- \Big) \, .
\ee
Here the (suitably rescaled) Pauli matrices $\sigma_3, \sigma_{\pm}$ are the generators of the Lie 
algebra $su(2)\simeq sl(2)$.

On $\R^4$ (and $\S^4$) the Hodge  operator $\ast_\theta$ acts as ``the undeformed one'' (remember that we do not deform the Dirac operator, i.e. the metric) and its action is explicitly found to be given by 
\be
\begin{array}{ll}
\ast_\theta \dd z_1 \dd z_2 =  \dd z_1 \dd z_2 & ~ \\
\ast_\theta \dd z_1 \dd z_1^* = \dd z_2 \dd z_2^*, & \ast_\theta \dd z_2 \dd z_2^* =  \dd z_1 \dd z_1^* ,  \\
\ast_\theta \dd z_1 \dd z_2^* = - \dd z_1 \dd z_2^*, & \ast_\theta \dd z_2 \dd z_1^* = - \dd z_2 \dd z_1^* .
\end{array}
\ee
When acting on \eqref{gcur} we get $\ast_\theta(u F_{(u)} u^*) = - u F_{(u)} u^*$, i.e. the curvature is antiself-dual.

As mentioned, when the deformation parameter  is set to zero, $\theta=0$, the gauge potential 
\eqref{gpot} is the BPST instanton \cite{BPST} of $\SU(2)$ Yang-Mills theory. 

\section{Quantum groups symmetries}
The noncommutative spheres $S^N_\Theta$ are  quantum homogeneous spaces of quantum orthogonal groups 
$\SO_\Theta(N+1)$, that is there is a coaction of  $\SO_\Theta(N+1)$ on the functions $A(S^N_\Theta)$ 
leaving the latter coinvariant \cite{V01,CD02}.  Dually, the symmetry can be realized \cite{S01} as the 
action of a universal envelopping algebra ${U_\Theta(so(N+1))}$.  In the present paper we shall describe in 
details the 4 dimensional case by giving the twisted symmetry action  of $\env$ on $\S^4$ which we then  lift 
to $\Sk^7$ in a way that the instanton potential decribed above is symmetric under the action. 

We recall that the eight roots of the Lie algebra $so(5)$ are two-component vectors $r=(r_1,r_2)$ of the 
form 
$r=(\pm1,\pm1), r=(0,\pm1), r=(\pm1,0)$.  There are corresponding generators $E_r$ of $so(5)$ toghether 
with two mutually commuting generators $H_1,H_2$ of the Cartan subalgebra. The Lie brackets are
\be\label{lie}
[H_1,H_2] = 0, \quad [H_j,E_r] = r_j E_r , \quad [E_{-r},E_{r}] = r_1 H_1 + r_2 H_2, \quad 
[E_{r},E_{r'}] = N_{r,r'} E_{r+r'}, 
\ee
 with $N_{r,r'}=0$ if $r+r'$ is not a root. The universal envelopping algebra $\oenv$ is the algebra 
 generated by elements $\{H_j, E_r\}$ modulo relations given by the previous Lie brackets\footnote{There 
 are additional Serre relations; they generate an ideal that needs to be quotiented out. This is not 
 problematic and we shall not dwell upon this point here.}. Then, the twisted universal envelopping algebra $\env$ is generated as above (i.e. one does not change the algebra structure) but is endowed with a twisted 
coproduct, 
\be
\Delta_\theta: \env \to \env \ot \env, 
\ee
which, on the generators 
$E_r$, $H_j$, reads
\bea\label{twdel} 
&& \Delta_\theta(E_r) = E_r \ot \lambda^{-r_1 H_2} + \lambda^{-r_2 H_1} \ot E_r , \nn\\
&& \Delta_\theta(H_j) = H_j  \ot \II + \II \ot H_j ,
\eea
with $\lambda=e^{2\pi \ii \theta}$ as before. This coproduct allows to represent $\env$ as an algebra 
of twisted ``differential operators''  (i.e. derivations of the algebras of functions) on both $\S^4$ and 
$\Sk^7$ as we shall see  below. With counit and antipode given by 
\bea\label{twhopf} 
&& \varepsilon(E_r) = \varepsilon(H_j) = 0, \nn\\
&& S(E_r) = - \lambda^{r_2 H_1}  E_ r\lambda^{r_1 H_2}, \quad S(H_j) = -H_j ,
\eea
 the algebra $\env$ becomes a Hopf algebra \cite{CP94}. At the classical value of the deformation 
parameter, $\theta=0$, one recovers the Hopf algebra structure of $\oenv$.

 We are ready for the representation of $\env$ on $\S^4$. For convenience, we introduce ``partial derivatives'', 
$\partial_\mu$ and 
 $\partial_\mu^*$ with the usual action on the generators of the algebra $A(\S^4)$: 
 $\partial_\mu(z_\nu)=\delta_{\mu\nu}$, $\partial_\mu(z_\nu^*)=0$, and 
 $\partial_\mu^*(z_\nu^*)=\delta_{\mu\nu}$, $\partial_\mu^*(z_\nu)=0$. Then, the action of $\env$ on 
$A(\S^4)$ is given by the following operators,
\begin{align}\label{act4}
H_1 &= z_1 \partial_1 - z_1^* \partial_1^* \, ,  & H_2 = z_2 \partial_2 - z_2^* \partial_2^*  \, , \nn\\
 E_{+1,+1} &= z_2 \partial_1^* - z_1 \partial_2^* \, ,  & E_{+1,-1} = z_2^* \partial_1^* - z_1 
\partial_2  \, , \nn\\
E_{+1,0} &= \frac{1}{\sqrt{2}}\, (2 z_0 \partial_1^* - z_1 \partial_0) \, ,  & E_{0,+1} = \frac{1}{\sqrt{2}}\, 
(2 z_0 \partial_2^* - z_2 \partial_0) \, , 
\end{align}
and $E_{-r}=(E_{r})^*$, with the obvious meaning of the adjoint.  These operators (not the partial 
derivatives!) are extended to the whole of $A(\S^4)$ as twisted derivations via the coproduct \eqref{twdel},
\bea\label{tder} 
&& E_r( a b ) = \Delta_\theta(E_r) (a \ot b) = E_r(a) \lambda^{-r_1 H_2}(b) + \lambda^{-r_2 H_1}(a) E_r(b) , \nn\\
&& H_j( a b ) = \Delta_\theta(H_j) (a \ot b) = H_j(a) b + a H_j(b) , 
\eea
for any two elements $a,b\in A(\S^4)$. Accordingly,  the partial derivatives $\partial_\mu$, $\partial_\mu^*$ 
will obey a twisted Leibniz rule for products of generators.  With these twisted rules, one readily 
checks compatibility with the commutation relations \eqref{s4t} of $A(\S^4)$. 

The representation of $\env$ on $\S^4$ given in \eqref{act4} is the fundamental vector representation.  
 When lifted to $\Sk^7$ one gets the fundamental spinor representation: as we see from the 
 quadratic relations among corresponding generators, as given in \eqref{sub}, the lifting amounts to 
take the ``square root'' representation.
 The action on $\env$ on $A(\Sk^7)$ is constructed by requiring twisted derivation properties via the 
 coproduct \eqref{tder} so as to reduce to the action \eqref{act4} on $A(\S^4)$ when using the defining 
 quadratic relations \eqref{sub}. The lifted action can be given as the action of twisted matrices 
$\Gamma$'s on the $\psi$'s,
\be\label{act7}
\psi_a \mapsto \sum_b \Gamma_{ab} \psi_b .
\ee
These matrices $\Gamma$ are the commutators of the twisted Dirac matrices in \eqref{tga}. By writing 
$\Gamma = \{H_j, E_r\}$, a long but straightforward computation shows that in the spinorial representation, they are given explicitly by
\begin{align}\label{tma}
& H_1 = \half\left( \begin{smallmatrix}
1  & & & \\
 & -1 &  &   \\
 &   & -1 &  \\
 &  & & 1 
\end{smallmatrix} \right), 
\qquad 
H_2 = \half\left( \begin{smallmatrix}
-1  & & & \\
 & 1 &  &   \\
 &   & -1 &  \\
 &  & & 1 
\end{smallmatrix} \right), \nn \\
& E_{+1,+1} =\begin{pmatrix} 
0 & 0 \\ 
0 & \begin{smallmatrix} 0 & -1 \\ 0 & 0 \end{smallmatrix} \\
\end{pmatrix}, 
\qquad 
E_{+1,-1} =\begin{pmatrix} 
\begin{smallmatrix} 0 & 0 \\ -\mu & 0 \end{smallmatrix} & 0 \\ 
0 & 0 \\
\end{pmatrix},  \nn \\
& E_{+1,0} = \frac{1}{\sqrt{2}} \begin{pmatrix} 
0 & \begin{smallmatrix} 0 & 0\\ 0 & -1 \end{smallmatrix} \\ 
\begin{smallmatrix} \mu  & 0 \\ 0 & 0 \end{smallmatrix} & 0 \\
\end{pmatrix}, 
\qquad
E_{0,+1} = \frac{1}{\sqrt{2}} \begin{pmatrix} 
0 & \begin{smallmatrix} 0 & \bar{\mu} \\ 0 & 0 \end{smallmatrix} \\ 
\begin{smallmatrix} 0  & 1 \\ 0 & 0 \end{smallmatrix} & 0 \\
\end{pmatrix}, 
\end{align}
and $E_{-r}=(E_{r})^*$ as before. As for the $\psi^*$'s, one finds instead that they transform as  $\psi^*_a \mapsto \sum_b (\wt{\Gamma})_{ab} \psi^*_b$, with 
\[
\wt{\Gamma} = \sigma \Gamma \sigma^{-1}, \qquad 
\sigma =\begin{pmatrix} 
\begin{smallmatrix} 0 & -1 \\ 1 & 0 \end{smallmatrix} & 0 \\ 
0 & \begin{smallmatrix} 0 & -1 \\ 1 & 0 \end{smallmatrix} \\
\end{pmatrix} ;
\] 
at $\mu=1$ (or $\theta=0$), one gets $\wt{\Gamma} = -\Gamma^*$. 
With the twisted rules \eqref{tder} for the action on products, one checks compatibility of the above 
action with the commutation relations \eqref{s7t} of $A(\Sk^7)$. 
Notice that the $\psi$'s and the $\psi^*$'s are not mixed by the action and each group transforms into itself.
It takes a little algebra to check that the instanton configuration, given in \eqref{co1fo} or \eqref{gpot} is 
invariant under the action of $\env$ in \eqref{act7} or \eqref{act4}.

 \bigskip
 
In fact, the polynomial algebra $A(\Sk^7)$ contains ALL representations of $\env$ while the polynomial 
subalgebra 
 $A(\S^4)$ contains only the vector representations. In a sense, $\Sk^7$ carries an action of 
 $U_\theta(spin(5))$ and dually a coaction of $\Spin_\theta(5)$, a fact that parallels the classical 
action of $\Spin(5)$ on $S^7\simeq \Spin(5) / \SU(2)$.
%
%

 From general considerations \cite{CP94}, the representation theory of $\env$ is the same as the one of 
 the undeformed counterpart $\oenv$, at least when  $\mu$ is not a root of unity; for the latter 
 case there may be additional representations.

There are two fundamental weights $W^1= \half (1,1)$ and $W^2=(1,0)$ which correspond to the fundamental 
spinor representation \eqref{act7} and vector representation \eqref{act4} respectively. 
Each representation is labelled by two integers $s,n$, being characterized by a highest weight $W=s W^1 + n W^2$ and has dimension 
\be
d(s,n) = (1+s)(1+n)(1+\frac{s+n}{2})(1+\frac{s+2n}{3}). 
\ee
The integer $s$ measures the ``spinorial content"  of the representation. A more usual (iso)-spin label 
 $I$, such that $2 I = s$, takes integer and half integer values and will be used later on. Since the 
Lie algebra structure is not changed, the quadratic Casimir operator is
\be\label{casi}
C = H_1^2 + H_2^2 + \sum_{r^+} (E_r E_{-r} + E_{-r} E_r) ,
\ee
with the sum over the positive roots. This operator  is a multiple of the identity in each representation, the 
factor 
being given by 
$C(s,n)=\half (s^2 + 2n^2 +2s n) +2s +3n$.

\section{The spin-Hall system on $\S^4$}

%
%

The model that we propose will parallel the one constructed in \cite{ZH01} on the classical sphere $S^4$. 
 The Hamiltonian of a ``single particle" moving on the sphere $\S^4$ and coupled to the gauge field 
$\eta=\eta_{(u=1)}$ in \eqref{gpot} is given by
\be\label{0ham}
H_{\eta} = -(\dd +\eta)^* (\dd + \eta) ,
\ee
with all physical constants being set to $1$. We take  the gauge potential $\eta$ in an arbitrary 
representation $I$ of the $su(2)$ algebra generated by  the Pauli matrices 
$\sigma_3, \sigma_\pm$. The isospin label $I$ takes integer  and half integer values and the dimension 
of this representation is $2I+1$ with corresponding Casimir  operator $C_{su(2)}$ taking the value 
$I(I+1)$. 

We can expand the covariant derivative, $D=\dd +\eta$,   on the basis of one forms $\dd 
z_\mu$,$\dd z_\mu^*$ as $D= \dd z_\mu D_\mu + \dd z_\mu^* D^*_\mu $.  Then, the Hamiltonian \eqref{0ham} 
takes the form,
\be\label{ham}
H_{\eta} = \wt{H}_1^2 + \wt{H}_2^2 + \sum_{r^+} (\wt{E}_r \wt{E}_{-r} + \wt{E}_{-r} \wt{E}_r)
\ee
where the operators $\wt{H}_j$ and  $\wt{E}_r$ are the same as in \eqref{act4} with the partial 
derivatives $\partial_\mu$, $\partial^*_\mu$  substituted by the covariant ones $D_\mu$, $D^*_\mu$ and the sum is over positive roots. 

Without the gauge potential the Hamiltonian  $H_{\eta=0}$ is just the Casimir operator \eqref{casi} of the 
twisted algebra $\env$. In a non vanishing gauge  field, one needs also the curvature $F=F_{(u=1)}$ of 
the connection. We shall then also expand the curvature $F$ given in \eqref{gcur} as a two form
\be
 F = F_{00} \dd z_0 \dd z_0 + \half F_{\varepsilon_{\mu}\mu, \varepsilon_{\nu}\nu} \dd z_{\varepsilon_{\mu}\mu} \dd 
z_{\varepsilon_{\nu}\nu}
\ee
 with $\varepsilon_{\mu}$ and $\varepsilon_{\nu}$ taking values $\pm 1$ (remember the root structure 
of $so(5)$) and with the convention that $\dd z_{-\mu}=\dd z_{\mu}^*$.
Some algebra shows that the operators
\be
 H_1=\wt{H}_1-F_{00}, \qquad H_2=\wt{H}_2-F_{00}, \qquad E_{\varepsilon_{\mu}\mu, \varepsilon_{\nu}\nu} = \wt{E}_{\varepsilon_{\mu}\mu, \varepsilon_{\nu}\nu} - 
F_{\varepsilon_{\mu}\mu, \varepsilon_{\nu}\nu} \, 
\ee
 close the commutation relations \eqref{lie} of the Lie algebra $so(5)$; the analogous classical result 
 was established in \cite{Y78b}. Now, the operators $F_{\varepsilon_{\mu}\mu, \varepsilon_{\nu}\nu}$ 
carry an isospin representation labelled by $I$. With this, one finds that the Hamiltonian 
 \eqref{ham} is given in terms of the Casimir operators $C$ and $C_{su(2)}$ (the latter coming from the instanton field 
strength) as
\be
H_{\eta} = C - 2C_{su(2)}.
\ee
The eigenvalues of this Hamiltonian are then given by the energies
\bea\label{energy}
E(I,n) &=& C(s=2I,n) - 2I(I+1)  \nn\\
&=& n^2+n(2I+3)+2I 
\eea
 with degeneracy $d(s=2I,n)$. The integer $n$ labels Landau levels and $I$, which plays the role of the 
magnetic flux, label the degeneracy in each Landau level.

 The ground state for a given $I$ is obtained when $n=0$; it has energy 
 \be
 E_0(I)=2I\ee
and its degeneration is 
\be
d_0(I)=d(s=2I,n=0)=\frac{1}{6}(1+2I)(2+2I)(3+2I).
\ee
  States with $n >0$ are separated 
from the ground state by a finite gap.
 Using the representation theory of $\env$ constructed above, it is straightforward to write down the 
 wave functions of the ground state. The spinor $\psi=(\psi_1,\dots,\psi_4)$ is an eigenfunction of 
 the Hamiltonian \eqref{ham} with $I=\half$, the fundamental spinorial representation, and from 
 \eqref{act7} and \eqref{tma} we see that the highest weight vector of this representation is $\psi_4$: 
 \be
 H_1(\psi_4)=\half=H_2(\psi_4).
 \ee
  Thus, a basis for the ground state, with is just the representation 
 with $s=2I$ and $n=0$, can be obtained by the corresponding highest weight vector, that is  
 $\Phi=(\psi_4)^{2I}$, by repeated action of the operators $E_r$ as twisted derivations. The results 
are $d_0(I)$ ($\mu$-symmetric) polynomials of degree $2I$ in $\psi_1,\dots,\psi_4$.

 With the local section \eqref{lose}, the ground state with ``orbital coordinates'' $\vec{z}=z_\mu$ and 
``isospin coordinates" $\vec{I}=(I_+,I_-,I_3)=(\sigma_+,\sigma_-,\sigma_3)$ is given by the wave functions,
\be\label{vac}
 \Phi(\vec{z},\vec{I};k_1,k_2,k_3,k_4) = Sym_\mu \Big((\psi_1)^{k_1} 
(\psi_2)^{k_2}(\psi_3)^{k_3}(\psi_4)^{k_4} \Big)\,
\ee
 with integer $k_1+k_2+k_3+k_4=2I$, and $Sym_\mu$ denotes a (tensor product) symmetrization which takes into account 
factors of the deformation parameter due to the commutation relations \eqref{s7t} of the $\psi$'s. 
 As said, the action of $\env$ on \eqref{vac} is constructed out of the fundamental spinorial 
representation \eqref{act7} with a repeated use of the twisted coaction \eqref{tder}.
 The previous are exact eigenstates of the Hamiltonian corresponding to the eigenvalue $E=2I$ with 
$d_0(I)$ degeneracy.

 There are also fractional excitations: they  are obtained from the highest weight vector 
$\Phi_q=(\Phi)^q=(\psi_4)^{2Iq}$, 
 with odd integer $q$. The corresponding  wave functions are fermionic eigenfunctions of the 
 Hamiltonian in the lowest Landau level and are indeed $\mu$-symmetric homogeneous polynomials of 
 degree $2 I q$ obtained, as before  by repeated action of the operators $E_r$ as twisted derivations. 
The degeneracy is now $d_0(qI)$ and the filling factor of the lowest Landau level is fractional,
\be
\nu = d_0(I) / d_0(qI) \sim q^{-3} + O(1/I), 
\ee
 
The large isospin limit $I \to \infty$ should be taken with care in order for the energy to remain finite 
\cite{ZH01}. One needs also to take $R \to\infty$ where $R$ is the  ``radius" of the sphere $\S^4$.
By inserting back physical quantities,  the energy eigenvalues are 
\[
E(I,n) = \frac{\hbar^2}{2m R^2}(n^2+n(2I+3)+2I).
\] 
The limit is taken  in such a manner that the ``magnetic length'' 
$\ell_0=\sqrt{2I/R^2}$ is constant.  In this limit, with a finite Landau level $n$, the energy is 
\be
E(I,n) \sim \frac{ \hbar^2 l_0^2 }{2m} (n+1).
\ee 
 
There are also ``orbital'' $\Orb$ symmetries; these are generated by the 
elements 
\be
L^+_a=\Big(\half(H_1+H_2), E_{+1,+1}, E_{-1,-1}\Big), \qquad 
L^-_a=\Big(\half(H_1-H_2), E_{+1,-1}, E_{-1,+1}\Big), 
\ee
with the label $a\in \{3,+,-\}$. One checks that the two sets $L^{\pm}_a$ close the $su(2)$ Lie algebra, having $so(4)=su(2)\op su(2)$. 
The presence of the instanton gauge potential will modify these generators to 
$\wt{L}^+_a=L^+_a$ and $\wt{L}^-_a=L^-_a+\sigma_a$ due to the commutation relations of the generators 
$\sigma_a$ which are here in a generic representation with isospin $I$. 

The representations of the corresponding envelopping algebra $\enor$ are determined by two spin labels
$(l^+,l^-)$ both taking integer and half integer values. It is straigthforward to check that 
the pair $(\psi_1,\psi_2)$ trasforms according to the representation $(0,\half)$ while the 
the pair $(\psi_4,\psi_3)$ trasforms according to the representation $(\half,0)$. Then, the ground 
states wave functions \eqref{vac} are in the representation 
\[
l^+=\half(k_3+k_4), \quad l^-=\half(k_1+k_2).
\]
Furthermore, they are also eigenstates 
of the operators $\wt{L}^+_3$ and $\wt{L}^-_3$ with eigenvalues given by
$l^+_3=k_4-k_3$ and $l^-_3=k_1-k_2$ respectively.

\section{Final remarks}

We have presented a gauge field theory on a noncommutative four sphere which gives explicit results. 
The gauge fields provides iso-spin degrees of freedom and there are symmetries under a quantum orthogonal group $\Ort$. The symmetries are enough to explicit diagonalize the Hamiltonian of excitations moving on the sphere $\S^4$ in the field of a noncommutative instanton. For the many body problem, there are both integer and half integer modes. The model is a generalization to noncommutative spaces of generalizations of spin-Hall models.

There are several open problems and questions starting with the physical meaning of the noncommutativity parameter. 
In the usual quantum Hall effect the filling factor plays the role of the deformation parameter and is directly accesible to experiments. In fact, the deformation parameter could be seen as the inverse of the magnetic field in which the carriers move. One can expect a similar interpretation for the model presented here. Also, one needs an analysis of the dependence on the  noncommutativity parameter $\mu$ of the ground states wave 
functions as well as of the quantum many body problem and of edge states.
Finally, there are potential applications to string theories and matrix models.

\vfill
\subsection*{Acknowledgments}
I am grateful to K.~Hasebe for providing me with a copy of the paper \cite{ZH01} and for useful remarks. W. van Suijlekom and J. V\'arilly have made a number suggestions. Finally, I thank the referees for helpful  comments.

\end{document}